\documentclass[aps,pra,preprint]{revtex4}

\usepackage{epsfig}
\usepackage{subfigure}

\begin{document}

\title{Non-Markovian Quantum State Diffusion: Application to Quantum Jumps in $^{24}$Mg$^+$}
\author{Joshua Wilkie and Ray Ng}
\affiliation{Department of Chemistry, Simon Fraser University, Burnaby, British Columbia V5A 1S6, Canada}

\date{\today}
 
\begin{abstract}

Non-Markovian quantum state diffusion (NMQSD) is an exact method for calculating the reduced density
matrix of an arbitrary subsystem interacting linearly with the radiation field. Applications 
of the theory have however been few due to the intractable nature of the variational-differential
NMQSD evolution equation. Recently, we argued that the variational-differential equation can be rewritten
as an integrodifferential equation which can be readily solved numerically. This manuscript provides
an explicit derivation of the modified equations. Applications to intermittent fluorescence in $^{24}$Mg$^+$
are discussed in detail. Earlier speculations that quantum jumps occur on all time scales are verified
on a picosecond timescale. We show that a plot of the probability density of the signal vs signal strength
shows the two characteristic peaks associated with the bright and dark manifolds, and that the ratio of
the areas under the peaks is 16 as observed experimentally. We also show that the shape of this distribution
is sensitive to bath memory, but has a mathematical form common to both the Markovian and non-Markovian cases.

\end{abstract}

\maketitle

\section{Introduction}

Like quantum state diffusion\cite{QSD}, non-Markovian quantum state diffusion (NMQSD)\cite{NMSD,NMSD2,NMSD3} was originally introduced as an 
exact computational 
method for finding the reduced density matrix of an arbitrary subsystem interacting linearly with
the radiation field. In practice the evolution equation proved impossible to solve except in a 
few cases where exact solutions were already known\cite{NMSD2}. Recently we reformulated NMQSD in terms of a solvable 
integrodifferential equation and demonstrated the use of the 
modified method by solving a number of example problems\cite{WN}. Here we present a detailed and general derivation of the
modified equations for an arbitrary number of coupling operators. We also apply the resulting theory to intermittent
fluorescence in driven $^{24}$Mg$^+$ which arises due to quantum jumps between bright and dark electronic states. 

Ion trap experiments have become an important subfield of quantum optics. Theory\cite{Theor,Theor2,Master2} and experiments\cite{OJump,Jump} on quantum jump phenomena in single ions are motivating further attempts to understand the measurement process\cite{Wise,Meas}. Such
ions have been used as models of quantum computers\cite{QC}, and there are many interesting possibilities when optical lattices
are employed\cite{Stein}. In many cases the subsystem of interest in these experiments can be modeled as a driven few level
system interacting linearly with the radiation field. NMQSD is an exact dynamical theory for such systems which should in principle
make theoretical analysis of such experiments a simple task. Unfortunately, standard formulations of NMQSD rely on a stochastic 
variational-differential equation (VDE) for dynamical evolution which cannot be solved - even numerically - outside of a few
special cases where solutions were already known. Some approximation schemes have recently been explored\cite{TYu,SY}, but exact 
numerical schemes remain an important goal.

In a recent manuscript we argued that NMQSD can be reformulated in terms of a solvable stochastic integrodifferential equation\cite{WN}.
We solved a number of example problems - some previously unsolvable - with the reformulated equations and found that accurate 
solutions to few level problems were readily obtainable. The efficiency of the method is partly a consequence of the availability 
of improved methods for solving stochastic differential equations (SDEs)\cite{SDE,COMM}. In this manuscript we apply the 
reformulated theory to $^{24}$Mg$^+$. Our results confirm speculation that quantum jumps occur on short time scales\cite{Jump}. We show that both Markovian and non-Markovian quantum state diffusion predict the two
peaked distribution seen in experiments. In addition both theories give the correct ratio for the areas under the peaks. We also
find that the mathematical lineshape functions are very similar for both theories even though the peaks appear qualitatively
different in the two cases.  

The derivation of the reformulated equations for NMQSD is outlined in section II. The equations and their numerical implementation
are discussed in section III. In section IV we construct a three level model for $^{24}$Mg$^+$. Numerical results for application
of NMQSD to the $^{24}$Mg$^+$ model are discussed in section V.

\section{Derivation of dynamical equation}

\subsection{Hermitian coupling}

For simplicity we will consider the simple special case where the total Hamiltonian is
\begin{eqnarray}
\hat{H}_{tot}&=&\hat{H}+\hat{x}\sum_{j=1}^mg_j (\hat{a}_{j}^{\dag}+\hat{a}_{j})+\sum_{j=1}^m\hbar \omega_j \hat{a}_j^{\dag}\hat{a}_j\nonumber \\
&=&\hat{H}+\hat{H}_c
\end{eqnarray}
and comment on the modifications necessary for more general Hamiltonians. We will also assume a bath temperature of 0 K and hence an initial state of the form
\begin{equation}
|\Psi_{tot}(0)\rangle =|\psi_0\rangle \otimes |0\rangle \dots \otimes  |0\rangle
\end{equation}
where $|0\rangle$ denotes the lowest eigenstate of $\hat{a}_j^{\dag}\hat{a}_j$. We will first show that the reduced
density 
\begin{equation}
\hat{\rho}(t)={\rm Tr}_{bath}\{ e^{-i\hat{H}_{tot}t/\hbar}|\Psi_{tot}(0)\rangle \langle \Psi_{tot}(0)|e^{i\hat{H}_{tot}t/\hbar}\}
\label{EQ1}
\end{equation}
can be rewritten as an average over diadics
\begin{equation}
\hat{\rho}(t)={\rm M}[|\psi_t\rangle\langle \psi_t|].
\end{equation}
To do this we will express the trace over bath modes as integrals over coherent states $|\alpha_j\rangle$,
\begin{equation}
\hat{a}_j|\alpha_j\rangle =\alpha_j |\alpha_j\rangle,
\end{equation}
for each oscillator. We do this by inserting closure relations 
\begin{equation}
\int d^2\alpha_j |\alpha_j\rangle \langle \alpha_j |=\hat{1}_j
\end{equation}
in Eq. (\ref{EQ1}), where $d^2\alpha_j=d {\rm Re}\alpha_j ~d {\rm Im}\alpha_j/\pi$. If $x,y$ denote eigenvalues of $\hat{x}$, and inserting closure relations $\int_{-\infty}^{\infty} dx'|x'\rangle\langle x'|=\hat{1}$ and $\int_{-\infty}^{\infty} dy'|y'\rangle\langle y'|=\hat{1}$, then it follows that matrix elements of the reduced density can be expressed as
\begin{eqnarray}
\langle x|\hat{\rho}(t)|y \rangle&&=\int_{-\infty}^{\infty} dx' \int_{-\infty}^{\infty} dy' \psi_0(x')\psi_0^*(y')\int d^2\alpha_1\dots \int d^2\alpha_m \nonumber \\
&&\langle x,\alpha_1,\dots, \alpha_m|e^{-i\hat{H}_{tot}t/\hbar}|x',0,\dots,0\rangle \langle y',0,\dots,0|e^{i\hat{H}_{tot}t/\hbar}|y,\alpha_1,\dots, \alpha_m\rangle.
\label{EQ2}
\end{eqnarray}
We now define $dt=t/N$ and use the Trotter product formula
\begin{equation}
e^{-i\hat{H}_{tot}t/\hbar}=\lim_{N\rightarrow \infty}e^{-i\hat{H}dt/\hbar}e^{-i\hat{H}_cdt/\hbar}\dots e^{-i\hat{H}dt/\hbar}e^{-i\hat{H}_cdt/\hbar}
\label{EE1}
\end{equation}
where there are now $2N$ factors inside the limit. Note that for more general Hamiltonians where there are many coupling operators $\hat{L}_k$ which couple to different modes of the bath it is necessary to use a appropriate generalization of the Trotter product formula to separate the subsystem Hamiltonian $\hat{H}$ and coupling operators $\hat{L}_k$ in separate factors. After this each coupling operator can be treated separately using techniques similar to those which follow.

Inserting $N-1$ closure relations for $x$ in Eq. (\ref{EE1}) and inserting the result into the matrix element $\langle x,\alpha_1,\dots, \alpha_m|e^{-i\hat{H}_{tot}t/\hbar}|x',0,\dots,0\rangle$ then gives 
\begin{eqnarray}
&&\langle x,\alpha_1,\dots, \alpha_m|e^{-i\hat{H}_{tot}t/\hbar}|x',0,\dots,0\rangle=\lim_{N\rightarrow\infty} \int_{-\infty}^{\infty}dx_1\dots \int_{-\infty}^{\infty}dx_{N-1} \nonumber \\
&&\langle x|e^{-i\hat{H}dt/\hbar}|x_{N-1}\rangle \dots \langle x_1|e^{-i\hat{H}dt/\hbar}|x'\rangle \nonumber \\
&&\prod_{j=1}^m \langle \alpha_j|e^{-i[x_{N-1}g_j(\hat{a}_{j}^{\dag}+\hat{a}_{j})+\hbar \omega_j \hat{a}_j^{\dag}\hat{a}_j]dt/\hbar} \dots e^{-i[x_{1}g_j(\hat{a}_{j}^{\dag}+\hat{a}_{j})+\hbar \omega_j \hat{a}_j^{\dag}\hat{a}_j]dt/\hbar}|0\rangle.
\end{eqnarray}
The factors in the coherent state matrix element can now be combined using 
\begin{equation}
e^{Adt}e^{Bdt}=e^{(A+B)dt+O(dt^2)}\label{COMB}
\end{equation}
and neglecting the $O(dt^2)$ terms. Introducing the usual path integral notation then gives
\begin{equation}
\langle x,\alpha_1,\dots, \alpha_m|e^{-i\hat{H}_{tot}t/\hbar}|x',0,\dots,0\rangle=\int_{x'}^x {\cal D}[x] e^{iS[x]/\hbar}\prod_{j=1}^m \langle \alpha_j|e^{-i[\int_0^t dt' x_{t'}g_j(\hat{a}_{j}^{\dag}+\hat{a}_{j})+\hbar \omega_j \hat{a}_j^{\dag}\hat{a}_jt]/\hbar}|0\rangle
\end{equation}
where $S[x]$ is the usual action.

The coherent state matrix elements can now be found by considering their dynamics. Defining
\begin{equation}
\psi_j(t,\alpha_j,\alpha_j^*)=\langle \alpha_j|e^{-i[\int_0^t dt' x_{t'}g_j(\hat{a}_{j}^{\dag}+\hat{a}_{j})+\hbar \omega_j \hat{a}_j^{\dag}\hat{a}_jt]/\hbar}|0\rangle
\end{equation}
and using the facts that $\langle \alpha_j |\hat{a}_j^{\dag}=\alpha^*_j \langle \alpha_j |$ and $\langle \alpha_j |\hat{a}_j=(\alpha_j/2+\partial/\partial \alpha^*_j)\langle \alpha_j |$ it then follows that $\psi_j(t,\alpha_j,\alpha_j^*)$ satisfies
\begin{equation}
d\psi_j(t,\alpha_j,\alpha_j^*)/dt=-(i/\hbar)\{x_tg_j(\alpha_j^*+\alpha_j/2+\partial/\partial \alpha_j^*)+\hbar \omega_j\alpha_j^*(\alpha_j/2+\partial/\partial \alpha_j^*)\}\psi_j(t,\alpha_j,\alpha_j^*)\label{EE2}
\end{equation}
with initial condition $\psi_j(0,\alpha_j,\alpha_j^*)=\langle \alpha_j|0\rangle=e^{-\alpha_j\alpha_j^*/2}$. Guessing a solution of the form
\begin{equation}
\psi_j(t,\alpha_j,\alpha_j^*)=\exp\{-\alpha_j\alpha_j^*/2+c_j(t)\alpha_j^*+d_j(t)\}\label{EAB}
\end{equation}
and substituting into Eq. (\ref{EE2}) one finds first order equations for unknowns $c_j(t)$ and $d_j(t)$ which
can be solved with initial conditions $c_j(0)=0$ and $d_j(0)=0$. The solutions are
\begin{eqnarray}
c_j(t)&=&-(i/\hbar)\int_0^tdt'~x_{t'}g_je^{-i\omega_j(t-t')}\nonumber\\
d_j(t)&=&-(1/\hbar^2)\int_0^tdt'\int_0^{t'}dt''~x_{t'}x_{t''}g_j^2e^{-i\omega_j(t'-t'')}\label{Prev}
\end{eqnarray}
and hence we may rewrite the matrix element in the form
\begin{eqnarray}
&&\langle x,\alpha_1,\dots, \alpha_m|e^{-i\hat{H}_{tot}t/\hbar}|x',0,\dots,0\rangle=\int_{x'}^x {\cal D}[x] e^{iS[x]/\hbar}\prod_{j=1}^m e^{-\alpha_j\alpha_j^*/2} \nonumber \\
&&e^{-(i/\hbar)\int_0^tdt'~x_{t'}g_je^{-i\omega_j(t-t')}\alpha_j^*}e^{-(1/\hbar^2)\int_0^tdt'\int_0^{t'}dt''~x_{t'}x_{t''}g_j^2e^{-i\omega_j(t'-t'')}}.
\end{eqnarray}
A similar formula can be found for $\langle y',0,\dots,0|e^{i\hat{H}_{tot}t/\hbar}|y,\alpha_1,\dots, \alpha_m\rangle$ and when both are substituted into Eq. (\ref{EQ2}) we obtain
\begin{eqnarray}
&&\langle x|\hat{\rho}(t)|y \rangle=\int_{-\infty}^{\infty} dx' \int_{-\infty}^{\infty} dy' \psi_0(x')\psi_0^*(y')\int_{x'}^x {\cal D}[x] \int_{y'}^y {\cal D}[y]  e^{iS[x]/\hbar}e^{-iS[y]/\hbar}\nonumber \\
&&\prod_{j=1}^m \int d^2\alpha_j~e^{-\alpha_j\alpha_j^*}e^{-(i/\hbar)\int_0^tdt'~x_{t'}g_je^{-i\omega_j(t-t')}\alpha_j^*}e^{(i/\hbar)\int_0^tdt'~y_{t'}g_je^{i\omega_j(t-t')}\alpha_j}\nonumber \\
&& e^{-(1/\hbar^2)\int_0^tdt'\int_0^{t'}dt''~x_{t'}x_{t''}g_j^2e^{-i\omega_j(t'-t'')}}
e^{-(1/\hbar^2)\int_0^tdt'\int_0^{t'}dt''~y_{t'}y_{t''}g_j^2e^{i\omega_j(t'-t'')}}.
\end{eqnarray}
The integrals over the real and imaginary parts of the $\alpha_j$ are now just Gaussian integrals which
can be performed analytically giving
\begin{eqnarray}
&&\langle x|\hat{\rho}(t)|y \rangle=\int_{-\infty}^{\infty} dx' \int_{-\infty}^{\infty} dy' \psi_0(x')\psi_0^*(y')\int_{x'}^x {\cal D}[x] \int_{y'}^y {\cal D}[y]  e^{iS[x]/\hbar}e^{-iS[y]/\hbar}\nonumber \\
&&e^{\int_0^tdt'\int_0^{t}dt''~x_{t'}y_{t''}\alpha^*(t',t'')}
e^{-\int_0^tdt'\int_0^{t'}dt''~x_{t'}x_{t''}\alpha(t',t'')}
e^{-\int_0^tdt'\int_0^{t'}dt''~y_{t'}y_{t''}\alpha^*(t',t'')}
\label{EE3}
\end{eqnarray}
where $\alpha(t,t')=(1/\hbar^2)\sum_{j=1}^mg_j^2e^{-i\omega_j(t-t')}$. 

The generalization to non-zero temperatures requires consideration of all $\langle x,\alpha_1,\dots, \alpha_m|e^{-i\hat{H}_{tot}t/\hbar}|x',n_1,\dots,n_m\rangle$ matrix elements (and their $y$-$y'$ counterparts) for all $n_1,\dots,n_m$ quantum numbers. The equation (\ref{EAB}) for $\psi_j(t,\alpha_j,\alpha_j^*)$ is unaltered but the initial condition is now $\frac{\alpha_j^{*n_j}}{\sqrt{n_j!}}e^{-\alpha_j\alpha_j^*/2}$ and so the correct ansatz is
\begin{equation}
\psi_j(t,\alpha_j,\alpha_j^*)=\frac{(\alpha^*_j+b_j(t))^{n_j}}{\sqrt{n_j!}}\exp\{-\alpha_j\alpha_j^*/2+c_j(t)\alpha_j^*+d_j(t)\}\label{EAB2}
\end{equation}
and the solutions for $b_j(t)$, $c_j(t)$ and $d_j(t)$ are 
\begin{eqnarray}
b_j(t)&=&-(i/\hbar)\int_0^tdt'~x_{t'}g_je^{i\omega_j(t-t')}\nonumber\\
c_j(t)&=&-(i/\hbar)\int_0^tdt'~x_{t'}g_je^{-i\omega_j(t-t')}\nonumber\\
d_j(t)&=&-(1/\hbar^2)\int_0^tdt'\int_0^{t'}dt''~x_{t'}x_{t''}g_j^2e^{-i\omega_j(t'-t'')}-in_j\omega_jt.
\end{eqnarray}
The correct thermal weight for mode $j$ is $e^{-\hbar\omega_jn_j/k_BT}(1-\exp(-\hbar\omega_j/k_BT))$ and introducing the notation
$w_j=\exp(-\hbar\omega_j/k_BT)$ we get
\begin{eqnarray}
&&\langle x|\hat{\rho}(t)|y \rangle=\int_{-\infty}^{\infty} dx' \int_{-\infty}^{\infty} dy' \psi_0(x')\psi_0^*(y')\int_{x'}^x {\cal D}[x] \int_{y'}^y {\cal D}[y]  e^{iS[x]/\hbar}e^{-iS[y]/\hbar}\nonumber \\
&&\prod_{j=1}^m \int d^2\alpha_j(1-w_j)e^{-\alpha_j\alpha_j^*}e^{-(i/\hbar)\int_0^tdt'~x_{t'}g_je^{-i\omega_j(t-t')}\alpha_j^*}e^{(i/\hbar)\int_0^tdt'~y_{t'}g_je^{i\omega_j(t-t')}\alpha_j}\nonumber \\
&& e^{-(1/\hbar^2)\int_0^tdt'\int_0^{t'}dt''~x_{t'}x_{t''}g_j^2e^{-i\omega_j(t'-t'')}}
e^{-(1/\hbar^2)\int_0^tdt'\int_0^{t'}dt''~y_{t'}y_{t''}g_j^2e^{i\omega_j(t'-t'')}}\nonumber\\
&&\sum_{n_j=0}^{\infty}\frac{e^{-\frac{\hbar n_j\omega_j}{k_BT}}}{n_j!}\{(\alpha_j+(i/\hbar)\int_0^tdt'y_{t'}g_j e^{-i\omega_j(t-t')}) (\alpha_j^*-(i/\hbar)\int_0^tdt'x_{t'}g_j e^{i\omega_j(t-t')})                 \}^{n_j}.
\end{eqnarray}
Performing the sum explicitly then gives
\begin{eqnarray}
&&\langle x|\hat{\rho}(t)|y \rangle=\int_{-\infty}^{\infty} dx' \int_{-\infty}^{\infty} dy' \psi_0(x')\psi_0^*(y')\int_{x'}^x {\cal D}[x] \int_{y'}^y {\cal D}[y] e^{iS[x]/\hbar}e^{-iS[y]/\hbar} \nonumber \\
&&\prod_{j=1}^m \int d^2\alpha_j(1-w_j)e^{-\alpha_j\alpha_j^*}e^{-(i/\hbar)\int_0^tdt'~x_{t'}g_je^{-i\omega_j(t-t')}\alpha_j^*}e^{(i/\hbar)\int_0^tdt'~y_{t'}g_je^{i\omega_j(t-t')}\alpha_j}\nonumber \\
&& e^{-(1/\hbar^2)\int_0^tdt'\int_0^{t'}dt''~x_{t'}x_{t''}g_j^2e^{-i\omega_j(t'-t'')}}
e^{-(1/\hbar^2)\int_0^tdt'\int_0^{t'}dt''~y_{t'}y_{t''}g_j^2e^{i\omega_j(t'-t'')}}\nonumber\\
&&e^{w_j(\alpha_j+(i/\hbar)\int_0^tdt'y_{t'}g_j e^{-i\omega_j(t-t')}) (\alpha_j^*-(i/\hbar)\int_0^tdt'x_{t'}g_j e^{i\omega_j(t-t')})}
\end{eqnarray}
or with some rearrangement
\begin{eqnarray}
&&\langle x|\hat{\rho}(t)|y \rangle=\int_{-\infty}^{\infty} dx' \int_{-\infty}^{\infty} dy' \psi_0(x')\psi_0^*(y')\int_{x'}^x {\cal D}[x] \int_{y'}^y {\cal D}[y] e^{iS[x]/\hbar}e^{-iS[y]/\hbar} \nonumber \\
&&\prod_{j=1}^m \int d^2\alpha_j(1-w_j)e^{-(1-w_j)\alpha_j\alpha_j^*}\nonumber \\
&&e^{[-(i/\hbar)\int_0^tdt'~x_{t'}g_je^{-i\omega_j(t-t')}+(i/\hbar)\int_0^tdt'~y_{t'}g_je^{-i\omega_j(t-t')}w_j]\alpha_j^*}\nonumber \\
&&e^{[(i/\hbar)\int_0^tdt'~y_{t'}g_je^{i\omega_j(t-t')}-(i/\hbar)\int_0^tdt'~x_{t'}g_je^{i\omega_j(t-t')}w_j]\alpha_j}\nonumber \\
&& e^{-(1/\hbar^2)\int_0^tdt'\int_0^{t'}dt''~x_{t'}x_{t''}g_j^2e^{-i\omega_j(t'-t'')}}
e^{-(1/\hbar^2)\int_0^tdt'\int_0^{t'}dt''~y_{t'}y_{t''}g_j^2e^{i\omega_j(t'-t'')}}\nonumber\\
&&e^{(1/\hbar^2)w_j\int_0^tdt'\int_0^tdt''x_{t'}y_{t''}g_j^2 e^{i\omega_j(t'-t'')}}.
\end{eqnarray}
Again the integrals over real and imaginary parts of $\alpha_j$ are Gaussian and can be done explicitly. The result is again Eq. (\ref{EE3}) where now 
$\alpha(t,t')=(1/\hbar^2)\sum_{j=1}^mg_j^2[\coth(\frac{\hbar\omega_j}{2k_BT}) \cos \omega_j(t-t')-i \sin \omega_j(t-t')]$. When the coupling operator is non-Hermitian both (\ref{EE3}) and the memory function must be modified\cite{NMSD2}.

Equation (\ref{EE3}) was first obtained by Feynman and Vernon\cite{Feyn}. The first exponential factor in this expression couples $x$ and $y$ which means the path integrals are coupled and must be performed simultaneously. This coupling can be eliminated at the expense of introducing a complex stochastic process $z_t$. Specifically, we use the identity
\begin{equation}
e^{\int_0^tdt'\int_0^{t}dt''~x_{t'}y_{t''}\alpha^*(t',t'')}={\rm M}[e^{\int_0^tdt'~(x_{t'}z_{t'}+y_{t'}z_{t'}^*)}],\label{EE4}
\end{equation}
first employed by Strunz\cite{NMSD}, where the mean over realizations of the noise is Gaussian
\begin{equation}
{\rm M}[F[z]]=\int {\cal D}[z]~F[z] ~Ne^{-\int_0^{\infty}dt'\int_0^{\infty}dt''~z_{t'}^*z_{t''}\beta (t',t'')}\label{EE5}
\end{equation}
and $\beta(t',t'')$ is the functional inverse of $\alpha(t,t')$ (i.e. $\int_0^{\infty}dt' ~\alpha(t,t')\beta(t',t'')=\delta(t-t'')$). Equation (\ref{EE4}) can be proved using (\ref{EE5}) by completing the square in the exponent.

Finally, inserting (\ref{EE4}) into Eq. (\ref{EE3}) we can rearrange terms so that
\begin{equation}
\langle x|\hat{\rho}(t)|y \rangle={\rm M}[\langle x|\psi_t\rangle \langle \psi_t|y\rangle ]
\end{equation}
where
\begin{equation}
\langle x|\psi_t\rangle=\int_{-\infty}^{\infty} dx'\psi_0(x')\int_{x'}^x {\cal D}[x]e^{iS[x]/\hbar}e^{\int_0^tdt'~x_{t'}z_{t'}} e^{-\int_0^tdt'\int_0^{t'}dt''~x_{t'}x_{t''}\alpha(t',t'')}.\label{DSG}
\end{equation}
To obtain a wave equation we differentiate with respect to time 
\begin{eqnarray}
&&d\langle x|\psi_t\rangle/dt=-(i/\hbar)\langle x|\hat{H}|\psi_t\rangle+\int_{-\infty}^{\infty} dx'\psi_0(x')\int_{x'}^x {\cal D}[x]e^{iS[x]/\hbar}\nonumber
\\
&&\{x_tz_te^{\int_0^tdt'~x_{t'}z_{t'}} e^{-\int_0^tdt'\int_0^{t'}dt''~x_{t'}x_{t''}\alpha(t',t'')}\nonumber \\
&&+e^{\int_0^tdt'~x_{t'}z_{t'}} (-x_t\int_0^{t}dt''~x_{t''}\alpha(t,t''))e^{-\int_0^tdt'\int_0^{t'}dt''~x_{t'}x_{t''}\alpha(t',t'')}\}\label{UGH}
\end{eqnarray}
and rewrite each term in terms of $\psi_t$. Noting that $x_t=x$ the second term presents no difficulties but the third does because of the delayed factor $x_{t''}$. Non-Markovian quantum state diffusion resolves this problem by introducing a 
variational derivative via the identity
\begin{equation}
x_{t''}e^{\int_0^tdt'~x_{t'}z_{t'}} =\frac{\delta}{\delta z_{t''}}e^{\int_0^tdt'~x_{t'}z_{t'}}.
\end{equation}
Thus, the linear NMQSD wave equation takes the form
\begin{equation}
\frac{d \psi_t}{dt}=-(i/\hbar)\hat{H}\psi_t+\hat{x}\psi_tz_t-\hat{x} \int_0^{t}dt'' \alpha(t,t'') \frac{\delta \psi_t}{\delta z_{t''}}.\label{EE6}
\end{equation}

Unfortunately, numerical methods for variational-differential equations like (\ref{EE6}) have not yet been developed. This means that few problems can be solved using (\ref{EE6}) and its generalizations\cite{NMSD2}. Recently, we have shown that there is a alternative way to formulate NMQSD in terms of a solvable integrodifferential equation\cite{WN}. The key idea is that Eq. (\ref{DSG}) can be rewritten in terms of a dynamical semigroup $\hat{U}(t,0)$ via $\langle x|\psi_t\rangle=\langle x|\hat{U}(t,0)|\psi_0\rangle$. To see that $\hat{U}(t,0)$ does represent a semigroup note that Eq. (\ref{DSG}) can be rewritten as
\begin{eqnarray}
&&\langle x|\psi_t\rangle=\int_{-\infty}^{\infty} dx'\psi_0(x')\int_{-\infty}^{\infty} dx_i
\int_{x_i}^x{\cal D}[x]e^{iS[x]/\hbar}e^{\int_{t_i}^tdt'~x_{t'}z_{t'}} e^{-\int_{t_i}^tdt'\int_0^{t'}dt''~x_{t'}x_{t''}\alpha(t',t'')}\nonumber \\
&&\int_{x'}^{x_i}{\cal D}[x]e^{iS[x]/\hbar}e^{\int_0^{t_i}dt'~x_{t'}z_{t'}} e^{-\int_0^{t_i}dt'\int_0^{t'}dt''~x_{t'}x_{t''}\alpha(t',t'')}
\label{DSGE}
\end{eqnarray}
for any intermediate time $t_i$. This means $\hat{U}(t,0)=\hat{U}(t,t_i)\hat{U}(t_i,0)$ and hence we have a semigroup. Applying this
to the second term in (\ref{UGH}) enables us to write
\begin{equation}
d\hat{U}(t,0)/dt=-(i/\hbar)\hat{H}\hat{U}(t,0)-i\hat{x}\hat{U}(t,0)z_t-\hat{x}\int_0^{t}dt'' \alpha(t,t'') \hat{U}(t,t'')\hat{x} \hat{U}(t'',0).
\end{equation}
All considerations above remain unchanged under $t\rightarrow -t$ and hence $\hat{U}(t,0)^{-1}$ exists. Using $\hat{U}(t,t'')=\hat{U}(t,0)\hat{U}^{-1}(t'',0)$ and changing notation to $U_t=U(t,0)$ we finally obtain
\begin{equation}
d\hat{U}_t/dt=-(i/\hbar)\hat{H}\hat{U}_t+\hat{x}\hat{U}_tz_t-\hat{x}\hat{U}_t\int_0^{t}dt'' \alpha(t,t'') \hat{U}_{t''}^{-1}\hat{x} \hat{U}_{t''}
\end{equation}
which is a closed integrodiffential equation for the propagator $\hat{U}_t$.

As in the case of NMQSD we must now find a norm-preserving version of the theory. This is accomplished
using the Girsonov transformation discussed at length in Ref. \cite{NMSD2}. The result is the evolution
equation
\begin{eqnarray}
&&d\hat{U}_t/dt=-(i/\hbar)\hat{H}\hat{U}_t-i(\hat{x}-\langle\hat{x} \rangle_t)\hat{U}_t(z_t+\int_0^{t}dt'' \alpha^*(t,t'')\langle\hat{x} \rangle_{t''})+C_t\hat{U}_t\nonumber \\
&&-(\hat{x}-\langle\hat{x} \rangle_t)  \hat{U}_t\int_0^{t}dt'' \alpha(t,t'') \hat{U}_{t''}^{-1}\hat{x} \hat{U}_{t''}
\end{eqnarray}
where the function $C_t$ is given by
\begin{equation}
C_t=\langle \psi_0|\hat{U}_t^{\dag}(\hat{x}-\langle\hat{x} \rangle_t)  \hat{U}_t\int_0^{t}dt'' \alpha(t,t'') \hat{U}_{t''}^{-1}\hat{x} \hat{U}_{t''}|\psi_0\rangle.
\end{equation}

\subsection{Non-Hermitian coupling}

The case where the coupling operator is non-Hermitian is slightly more complicated, although the general argument is similar. Consider
the Hamiltonian
\begin{eqnarray}
\hat{H}_{tot}&=&\hat{H}+\sum_{j=1}^mg_j (\hat{L}\hat{a}_{j}^{\dag}+\hat{L}^{\dag}\hat{a}_{j})+\sum_{j=1}^m\hbar \omega_j \hat{a}_j^{\dag}\hat{a}_j\nonumber \\
&=&\hat{H}+\hat{K}+\hat{K}^{\dag}
\end{eqnarray}
where $\hat{K}=\sum_{j=1}^mg_j \hat{L}\hat{a}_{j}^{\dag}+(1/2)\sum_{j=1}^m\hbar \omega_j \hat{a}_j^{\dag}\hat{a}_j$. Now suppose that we can find a complete or over-complete eigenbasis for $\hat{L}$ such that $\hat{L}|\lambda\rangle =\lambda |\lambda\rangle $ and $\int d^2\lambda |\lambda\rangle\langle\lambda |=1$. If $|\mu\rangle$ and $|\nu\rangle$ are two such states then
\begin{eqnarray}
&&\langle \mu|\hat{\rho}(t)|\nu\rangle  = \int d^2\mu'\int d^2\nu' \langle \mu'|\psi_0\rangle\langle \psi_0| \nu'\rangle \int d^2\alpha_1\dots \int d^2\alpha_m \nonumber \\
&& \sum_{n_1=0}^{\infty}\dots \sum_{n_m=0}^{\infty}\prod_{j=1}^m(1-w_j)w_j^{n_j} \nonumber \\
&&\langle \mu,\alpha_1,\dots ,\alpha_m|e^{-i\hat{H}_{tot}t/\hbar}|\mu',n_1,\dots , n_m\rangle \langle \nu',n_1,\dots ,n_m|e^{i\hat{H}_{tot}t/\hbar}|\mu,\alpha_1,\dots , \alpha_m\rangle 
\end{eqnarray}
where $w_j=e^{-\hbar\omega_{j}/k_BT}$.
Using the Trotter product formula
\begin{equation}
e^{-i\hat{H}_{tot}t/\hbar}=\lim_{N\rightarrow \infty}e^{-i\hat{H}dt/\hbar}e^{-i\hat{K}dt/\hbar}e^{-i\hat{K}^{\dag}dt/\hbar}\dots e^{-i\hat{H}dt/\hbar}e^{-i\hat{K}dt/\hbar}e^{-i\hat{K}^{\dag}dt/\hbar}
\end{equation}
and inserting closure relations between the $\hat{K}$ and $\hat{K}^{\dag}$ factors then gives
\begin{eqnarray}
&&\langle \mu,\alpha_1,\dots ,\alpha_m|e^{-i\hat{H}_{tot}t/\hbar}|\mu',n_1,\dots , n_m\rangle=\lim_{N\rightarrow\infty}\int d^2\mu_1\dots \int d^2\mu_{N-1}\nonumber \\
&&\langle \mu|e^{-i\hat{H}dt/\hbar}|\mu_{N-1}\rangle \dots \langle \mu_2|e^{-i\hat{H}dt/\hbar}|\mu_{1}\rangle \langle \mu_1|\mu'\rangle \nonumber \\
&& \prod_{j=1}^m \langle \alpha_j|e^{-(i/\hbar)[g_j\mu_{N-1}\hat{a}_j^{\dag}+(1/2)\hbar \omega_j \hat{a}_j^{\dag}\hat{a}_j]dt/\hbar}e^{-(i/\hbar)[g_j\mu_{N-1}^*\hat{a}_j^{\dag}+(1/2)\hbar \omega_j \hat{a}_j^{\dag}\hat{a}_j]dt/\hbar} \nonumber \\
&& \dots  e^{-(i/\hbar)[g_j\mu_{1}\hat{a}_j^{\dag}+(1/2)\hbar \omega_j \hat{a}_j^{\dag}\hat{a}_j]dt/\hbar}  e^{-(i/\hbar)[g_j\mu_{1}^*\hat{a}_j^{\dag}+(1/2)\hbar \omega_j \hat{a}_j^{\dag}\hat{a}_j]dt/\hbar}|n_j\rangle
\end{eqnarray}
and combining terms using Eq. (\ref{COMB}) and introducing path integral notation gives
\begin{eqnarray}
&&\langle \mu,\alpha_1,\dots ,\alpha_m|e^{-i\hat{H}_{tot}t/\hbar}|\mu',n_1,\dots , n_m\rangle=\int_{\mu'}^{\mu} {\cal D}[\mu] ~e^{iS[\mu]/\hbar}\nonumber \\
&&\prod_{j=1}^m \langle \alpha_j|e^{-(i/\hbar)[g_j(\int_0^tdt'\mu_{t'}\hat{a}_j^{\dag}+\int_0^tdt'\mu_{t'}^*\hat{a}_j)+\hbar \omega_j \hat{a}_j^{\dag}\hat{a}_jt]}|n_j\rangle.
\end{eqnarray}
The recipe for dealing with multiple coupling operators is a straightforward generalization of this result since each has its own oscillator bath.

Defining $\psi_j$ again via
\begin{equation} 
\psi_j(t,\alpha_j,\alpha_j^*)= \langle \alpha_j|e^{-(i/\hbar)[g_j(\int_0^tdt'\mu_{t'}\hat{a}_j^{\dag}+\int_0^tdt'\mu_{t'}^*\hat{a}_j)+\hbar \omega_j \hat{a}_j^{\dag}\hat{a}_jt]}|n_j\rangle
\end{equation}
and differentiating then gives
\begin{equation}
d\psi_j(t,\alpha_j,\alpha_j^*)/dt=-(i/\hbar)\{g_j(\mu_t\alpha_j^*+\mu_t^*(\alpha_j/2+\partial/\partial \alpha_j^*))+\hbar \omega_j\alpha_j^*(\alpha_j/2+\partial/\partial \alpha_j^*)\}\psi_j(t,\alpha_j,\alpha_j^*).
\end{equation}
The ansatz (\ref{EAB2}) works here also and consistency then requires
\begin{eqnarray}
b_j(t)&=&-(i/\hbar)\int_0^tdt'~\mu_{t'}^*g_je^{i\omega_j(t-t')}\nonumber\\
c_j(t)&=&-(i/\hbar)\int_0^tdt'~\mu_{t'}g_je^{-i\omega_j(t-t')}\nonumber\\
d_j(t)&=&-(1/\hbar^2)\int_0^tdt'\int_0^{t'}dt''~\mu_{t'}^*\mu_{t''}g_j^2e^{-i\omega_j(t'-t'')}-in_j\omega_jt.
\end{eqnarray}
Substituting this back into the subsystem density matrix element, performing the sum over the $n_j$ for each $j$, and then performing the integrals over $\alpha_j$ gives
\begin{eqnarray}
&&\langle \mu|\hat{\rho}(t)|\nu\rangle  = \int d^2\mu'\int d^2\nu' \langle \mu'|\psi_0\rangle\langle \psi_0| \nu'\rangle \int_{\mu'}^{\mu} {\cal D}[\mu] \int_{\nu'}^{\nu} {\cal D}[\nu] ~e^{iS[\mu]/\hbar} ~e^{-iS[\nu]/\hbar}\nonumber \\
&&e^{\int_0^tdt'\int_0^{t}dt'' \mu_{t'}\nu_{t''}^*\alpha^{-*}(t',t'')+\int_0^tdt'\int_0^{t}dt'' \mu_{t'}^*\nu_{t''}\alpha^{+*}(t',t'')}
e^{-\int_0^tdt'\int_0^{t'}dt'' \mu_{t'}^*\mu_{t''}\alpha^-(t',t'')-\int_0^tdt'\int_0^{t'}dt'' \mu_{t'}\mu_{t''}^*\alpha^+(t',t'')}\nonumber \\
&&
e^{-\int_0^tdt'\int_0^{t'}dt'' \nu_{t'}\nu_{t''}^{*}\alpha^{-*}(t',t'')-\int_0^tdt'\int_0^{t'}dt'' \nu_{t'}^*\nu_{t''}\alpha^{+*}(t',t'')}
\end{eqnarray}
where $\alpha^{-}(t',t'')=\sum_{j=1}^mg_j^2\frac{1}{1-w_j}e^{-i\omega_j(t'-t'')}$ and $\alpha^{+}(t',t'')=\sum_{j=1}^mg_j^2\frac{w_j}{1-w_j}e^{i\omega_j(t'-t'')}$. Now we introduce two independent complex noises $z_t^-$ and $z_t^+$ and unravel the coupled terms via
\begin{eqnarray}
&&e^{\int_0^tdt'\int_0^{t}dt'' \mu_{t'}\nu_{t''}^*\alpha^{-*}(t',t'')}={\rm M}[e^{\int_0^tdt'~(\mu_{t'}z_{t'}^{-}+\nu_{t'}^*z_{t'}^{-*})}]\nonumber \\
&&e^{\int_0^tdt'\int_0^{t}dt'' \mu_{t'}^*\nu_{t''}\alpha^{+*}(t',t'')}={\rm M}[e^{\int_0^tdt'~(\mu_{t'}^*z_{t'}^{+}+\nu_{t'}z_{t'}^{+*})}],
\end{eqnarray}
where $M[~]$ denotes the Gaussian average over both noises. It now follows that 
\begin{equation}
\langle \mu|\hat{\rho}(t)|\nu \rangle={\rm M}[\langle \mu|\psi_t\rangle \langle \psi_t|\nu \rangle ]
\end{equation}
where 
\begin{eqnarray}
\langle \mu|\psi_t\rangle =\int d^2\mu'\langle \mu'|\psi_0\rangle\int_{\mu'}^{\mu} {\cal D}[\mu] ~e^{iS[\mu]/\hbar}e^{\int_0^tdt'~(\mu_{t'}z_{t'}^{-}+\mu_{t'}^*z_{t'}^{+})}\nonumber \\
e^{-\int_0^tdt'\int_0^{t'}dt'' \mu_{t'}^*\mu_{t''}\alpha^-(t',t'')-\int_0^tdt'\int_0^{t'}dt'' \mu_{t'}\mu_{t''}^*\alpha^+(t',t'')}.
\end{eqnarray}
This path integral representation of the wavefunction is also of dynamical semigroup form and hence we can introduce a propagator $U_t$ via $\langle \mu|\psi_t\rangle=\langle \mu|U_t|\psi_0\rangle$ which then obeys 
\begin{eqnarray}
&&d\hat{U}_t/dt=-(i/\hbar)\hat{H}\hat{U}_t+\hat{L}\hat{U}_tz_t^-+\hat{L}^{\dag}\hat{U}_tz_t^+\nonumber \\
&&-\hat{L}^{\dag}\hat{U}_t\int_0^{t}dt'' \alpha^-(t,t'') \hat{U}_{t''}^{-1}\hat{L} \hat{U}_{t''}-\hat{L}\hat{U}_t\int_0^{t}dt'' \alpha^+(t,t'') \hat{U}_{t''}^{-1}\hat{L}^{\dag} \hat{U}_{t''}
\end{eqnarray}
and this can again be recast in a norm preserving form. 

\subsection{Multiple non-Hermitian couplings}

Finally, consider the case where we have $p$ non-Hermitian coupling operators and non-zero temperature. Consider a Hamiltonian
of the form
\begin{equation}
\hat{H}_{tot}=\hat{H}+\sum_{l=1}^p(\hat{K}_l+\hat{K}^{\dag}_l)
\end{equation}
where $\hat{K}_l=\sum_{j=1}^{m_l}g_j^l \hat{L}_l\hat{a}_{j,l}^{\dag}+(1/2)\sum_{j=1}^{m_l}\hbar \omega_{j,l} \hat{a}_{j,l}^{\dag}\hat{a}_{j,l}$
where a different subset of bath operators couples to each system operator $\hat{L}_l$. Now assume that each $\hat{L}_l$ has a 
complete or overcomplete eigenbasis $\hat{L}_l|\lambda_l\rangle =\lambda_l |\lambda_l\rangle $ and $\int d^2\lambda_l |\lambda_l\rangle\langle\lambda_l |=1$. Pick eigenstates $|\mu\rangle$ and $|\nu\rangle$ of $\hat{L}_1$, so that we may write
\begin{eqnarray}
&&\langle \mu|\hat{\rho}(t)|\nu\rangle  = \int d^2\mu'\int d^2\nu' \langle \mu'|\psi_0\rangle\langle \psi_0| \nu'\rangle \int d^2\alpha_1^1\dots \int d^2\alpha_{m_p}^p \nonumber \\
&& \prod_{l=1}^p\sum_{n_1^l=0}^{\infty}\dots \sum_{n_{m_l}^l=0}^{\infty}\prod_{j=1}^{m_l}(1-w_{j,l})w_{j,l}^{n_j^l} \nonumber \\
&&\langle \mu,\alpha_1^1,\dots ,\alpha_{m_p}^p|e^{-iH_{tot}t/\hbar}|\mu',n_1^1,\dots , n_{m_p}^p\rangle \langle \nu',n_1^1,\dots ,n_{m_p}^p|e^{iH_{tot}t/\hbar}|\mu,\alpha_1^1,\dots , \alpha_{m_p}^p\rangle \label{Multi}
\end{eqnarray}
where $w_{j,l}=e^{-\hbar\omega_{j,l}/k_BT}$. Now we employ a Trotter product formula 
\begin{eqnarray}
&&e^{-i\hat{H}_{tot}t/\hbar}=\lim_{N\rightarrow \infty}e^{-i\hat{H}dt/\hbar}e^{-i\hat{K}_1dt/\hbar}e^{-i\hat{K}^{\dag}_1dt/\hbar}\dots 
e^{-i\hat{K}_pdt/\hbar}e^{-i\hat{K}^{\dag}_pdt/\hbar} \nonumber \\
&&\dots e^{-i\hat{H}dt/\hbar}e^{-i\hat{K}_1dt/\hbar}e^{-i\hat{K}^{\dag}_1dt/\hbar}\dots e^{-i\hat{K}_pdt/\hbar}e^{-i\hat{K}^{\dag}_pdt/\hbar}
\end{eqnarray}
in which the $p+1$ factors are repeated $N$ times. Now we insert closure relations between each $e^{-i\hat{K}_idt/\hbar}e^{-i\hat{K}^{\dag}_idt/\hbar}$ pair. This results in a sort of overcomplete path integral representation in which
\begin{eqnarray}
&&\langle \mu,\alpha_1^1,\dots ,\alpha_{m_p}^p|e^{-i\hat{H}_{tot}t/\hbar}|\mu',n_1^1,\dots , n_{m_p}^p\rangle=\int_{\mu'}^{\mu} {\cal D}[\mu^1]\int {\cal D}[\mu^2]\dots \int {\cal D}[\mu^p] ~e^{iS[\mu^1,\dots,\mu^p]/\hbar}\nonumber \\
&&\prod_{l=1}^{p} \prod_{j=1}^{m_l} \langle \alpha_j^l|e^{-(i/\hbar)[g_j^l(\int_0^tdt'\mu_{t'}^l\hat{a}_j^{l\dag}+\int_0^tdt'\mu_{t'}^{l*}\hat{a}_j^l)+\hbar \omega_{j,l} \hat{a}_j^{l\dag}\hat{a}_j^lt]}|n_j\rangle.
\end{eqnarray}
Now, each factor can be handled as in the previous subsection. When the results are substituted into Eq. (\ref{Multi}) one obtains
\begin{eqnarray}
&&\langle \mu|\hat{\rho}(t)|\nu\rangle  = \int d^2\mu'\int d^2\nu' \langle \mu'|\psi_0\rangle\langle \psi_0| \nu'\rangle \int_{\mu'}^{\mu} {\cal D}[\mu^1]\int {\cal D}[\mu^2]\dots \int {\cal D}[\mu^p]\nonumber \\
&&\int_{\nu'}^{\nu} {\cal D}[\nu^1]\int {\cal D}[\nu^2]\dots \int {\cal D}[\nu^p] ~e^{iS[\mu^1,\dots,\mu^p]/\hbar}~e^{-iS[\nu^1,\dots,\nu^p]/\hbar}\nonumber \\
&&\prod_{l=1}^pe^{\int_0^tdt'\int_0^{t}dt'' \mu_{t'}^l\nu_{t''}^{l*}\alpha^{l-*}(t',t'')+\int_0^tdt'\int_0^{t}dt'' \mu_{t'}^{l*}\nu_{t''}^l\alpha^{l+*}(t',t'')}
e^{-\int_0^tdt'\int_0^{t'}dt'' \mu_{t'}^{l*}\mu_{t''}^l\alpha^{l-}(t',t'')-\int_0^tdt'\int_0^{t'}dt'' \mu_{t'}^l\mu_{t''}^{l*}\alpha^{l+}(t',t'')}\nonumber \\
&&
e^{-\int_0^tdt'\int_0^{t'}dt'' \nu_{t'}^l\nu_{t''}^{l*}\alpha^{l-*}(t',t'')-\int_0^tdt'\int_0^{t'}dt'' \nu_{t'}^{l*}\nu_{t''}^l\alpha^{l+*}(t',t'')}
\end{eqnarray}
from which the stochastic wave function can be obtained by introducing two independent complex noises for each of the $p$ coupling operators. The equation for the propagator can then be obtained along the lines followed in the previous subsections.

\section{General dynamical equations}

Consider a subsystem-bath model with multiple operators $\hat{L}_k$ interacting with different subsets of the radiation field
\begin{equation}
\hat{H}_{tot}=\hat{H}+\sum_{k=1}^{n}\sum_{j=1}^{m_k}g_{j}^k (\hat{L}_k \hat{a}_{j,k}^{\dag}+\hat{L}^{\dag}_k\hat{a}_{j,k})+\sum_{k=1}^n\sum_{j=1}^{m_k}\hbar \omega_{j,k} \hat{a}_{j,k}^{\dag}\hat{a}_{j,k}
\end{equation}
where $\hat{H}$ is the subsystem Hamiltonian, $\hat{L}_k$ are system coupling operators, and $g_{j}^k$ is a coupling
constant for an oscillator mode of frequency $\omega_{j,k}$ of manifold $k$. 
In the norm-preserving formulation of NMQSD at zero temperature the evolution of the state vector $\psi_t$ is governed by the VDE
\begin{eqnarray}
\frac{d\psi_t}{dt}&=&-(i/\hbar)\hat{H} ~\psi_t +\sum_{k=1}^n\tilde{z}_t^k (\hat{L}_k-\langle \hat{L}_k\rangle_t) ~\psi_t\nonumber \\
& -&\sum_{k=1}^n\int_0^tds~\alpha^k (t,s) ~(\hat{L}^{\dag}_k-\langle \hat{L}^{\dag}_k\rangle_t) ~\frac{\delta \psi_t}{\delta \tilde{z}_s^k}\nonumber \\
&+&\sum_{k=1}^n\int_0^tds~\alpha^k (t,s) ~\langle \psi_t|(\hat{L}^{\dag}_k-\langle \hat{L}^{\dag}_k\rangle_t) |\frac{\delta \psi_t}{\delta \tilde{z}_s^k}\rangle ~\psi_t\label{EQP1}
\end{eqnarray}
where $\tilde{z}_t^k=z_t^k+\int_0^tds~\alpha^{k*}(t,s)\langle \hat{L}^{\dag}_k\rangle_s$ and $z_t^k$ is a complex colored noise\cite{Cnoise} for manifold $k$ with correlation function $\alpha^k(t,s)=M[z_t^{k*}z_s^k]= (1/\hbar^2)\sum_{j=1}^mg_j^2e^{-i\omega_j(t-s)}$. 
The notation $\langle \hat{L}_k\rangle_t$ denotes the quantum expectation $\langle \psi_t|\hat{L}_k|\psi_t\rangle$ while $M[\dots ]$ denotes an average over different realizations of the noises. The exact reduced density matrix $\rho_t$ of the 
subsystem is given as an average of diadics via $\rho_t=M[|\psi_t\rangle \langle \psi_t|]$.

When we reformulate the theory in terms of a propagator $\psi_t=\hat{U}_t\psi_0$ then we obtain a closed set of equations
\begin{eqnarray}
\frac{d\hat{U}_t}{dt}&=&-(i/\hbar)\hat{H} \hat{U}_t +\sum_{k=1}^n ~(\hat{L}_k-\langle \hat{L}_k\rangle_t) ~\hat{U}_t ~(z_t^k+\int_0^tds~\alpha^{k*}(t,s)\langle \hat{L}^{\dag}_k\rangle_s)+C_t\hat{U}_t\nonumber \\
&-&\sum_{k=1}^n(\hat{L}^{\dag}_k-\langle \hat{L}^{\dag}_k\rangle_t)~\hat{U}_t\int_0^tds~\alpha^k (t,s) \hat{U}_s^{-1}\hat{L}_k\hat{U}_s\nonumber \\
\frac{d \hat{U}_t^{-1}}{dt}&=&-\hat{U}_t^{-1} \frac{d\hat{U}_t}{dt} \hat{U}_t^{-1}\label{GEQ5}
\end{eqnarray}
where 
\begin{eqnarray}
C_t=\sum_{k=1}^n\langle \psi_0|\hat{U}_t^{\dag}(\hat{L}^{\dag}_k-\langle \hat{L}^{\dag}_k\rangle_t)\hat{U}_t\int_0^tds~\alpha^k (t,s) \hat{U}_s^{-1}\hat{L}_k\hat{U}_s|\psi_0\rangle
\end{eqnarray}
depends on the initial state $\psi_0$. 

For non-zero temperatures and non-Hermitian coupling operators the equations are
\begin{eqnarray}
\frac{d\hat{U}_t}{dt}&=&-(i/\hbar)\hat{H} \hat{U}_t +\sum_{k=1}^n ~(\hat{L}_k-\langle \hat{L}_k\rangle_t) ~\hat{U}_t ~(z_t^{k-}+\int_0^tds~\alpha^{k-*}(t,s)\langle \hat{L}^{\dag}_k\rangle_s)\nonumber \\
&&+\sum_{k=1}^n ~(\hat{L}_k^{\dag}-\langle \hat{L}_k^{\dag}\rangle_t) ~\hat{U}_t ~(z_t^{k+}+\int_0^tds~\alpha^{k+*}(t,s)\langle \hat{L}_k\rangle_s)+C_t\hat{U}_t\nonumber \\
&&-\sum_{k=1}^n(\hat{L}^{\dag}_k-\langle \hat{L}^{\dag}_k\rangle_t)~\hat{U}_t\int_0^tds~\alpha^{k-} (t,s) \hat{U}_s^{-1}\hat{L}_k\hat{U}_s\nonumber \\
&&-\sum_{k=1}^n(\hat{L}_k-\langle \hat{L}_k\rangle_t)~\hat{U}_t\int_0^tds~\alpha^{k+} (t,s) \hat{U}_s^{-1}\hat{L}_k^{\dag}\hat{U}_s\nonumber \\
&&\frac{d \hat{U}_t^{-1}}{dt}=-\hat{U}_t^{-1} \frac{d\hat{U}_t}{dt} \hat{U}_t^{-1}
\end{eqnarray}
where 
\begin{eqnarray}
&&C_t=\sum_{k=1}^n\langle \psi_0|\hat{U}_t^{\dag}(\hat{L}^{\dag}_k-\langle \hat{L}^{\dag}_k\rangle_t)\hat{U}_t\int_0^tds~\alpha^{k-} (t,s) \hat{U}_s^{-1}\hat{L}_k\hat{U}_s|\psi_0\rangle\nonumber \\
&&+\sum_{k=1}^n\langle \psi_0|\hat{U}_t^{\dag}(\hat{L}_k-\langle \hat{L}_k\rangle_t)\hat{U}_t\int_0^tds~\alpha^{k+} (t,s) \hat{U}_s^{-1}\hat{L}_k^{\dag}\hat{U}_s|\psi_0\rangle
\end{eqnarray}

The most efficient way of solving these equations depends on the properties of the memory functions. We will
assume that the memory functions consist of a few terms of exponential form, i.e.,
\begin{equation}
\alpha^k (t,s)=\sum_{j=1}^{m_k} A_{j,k} e^{-\gamma_{j,k} |t-s|} e^{-i\omega_{j,k}(t-s)}\label{MEMORY}
\end{equation}
where $A_{j,k}$ and $\gamma_{j,k}$ are positive numbers. The terms do not in general correspond to
physical bath oscillator modes. Instead the expansion can be viewed as a best fit to the memory function, 
obtained by nonlinear least squares or other techniques. In many cases the number of required terms can be quite small. The case where the memory function cannot be represented this way is considered elsewhere\cite{WN}. Defining operators $\hat{V}_{t}^{j,k}=\int_0^t ds~ A_{j,k} e^{-\gamma_{j,k} (t-s)} e^{-i\omega_{j,k}(t-s)} U_s^{-1}\hat{L}_kU_s$ then Eqs. (\ref{GEQ5}) become
\begin{eqnarray}
\frac{d\hat{U}_t}{dt}&=&-(i/\hbar)\hat{H} \hat{U}_t +\sum_{k=1}^n ~(\hat{L}_k-\langle \hat{L}_k\rangle_t) ~\hat{U}_t ~(z_t^k+\sum_{j=1}^{m_k}y_t^{j,k})+C_t\hat{U}_t\nonumber \\
&-&\sum_{k=1}^n(\hat{L}^{\dag}_k-\langle \hat{L}^{\dag}_k\rangle_t)~\hat{U}_t\sum_{j=1}^{m_k}\hat{V}_t^{j,k}\nonumber \\
\frac{d \hat{U}_t^{-1}}{dt}&=&-\hat{U}_t^{-1} \frac{d\hat{U}_t}{dt} \hat{U}_t^{-1}\nonumber \\
\frac{d \hat{V}_{t}^{j,k}}{dt}&=&-(\gamma_{j,k}+i\omega_{j,k})\hat{V}_{t}^{j,k}+A_{j,k}\hat{U}_t^{-1}\hat{L}_k\hat{U}_t\nonumber \\
\frac{dy_{t}^{j,k}}{dt}&=&-(\gamma_{j,k}-i\omega_{j,k})y_{t}^{j,k}+A_{j,k}\langle \hat{L}^{\dag}_k\rangle_t\label{EEs}
\end{eqnarray}
where $y_{t}^{j,k}=\int_0^tds A_{j,k}e^{-\gamma_{j,k} (t-s)} e^{i\omega_{j,k}(t-s)} \langle \hat{L}^{\dag}_k\rangle_s$. Colored noises $z_t^k=\sum_{j=1}^{m_k}\xi_t^{j,k}$ can be generated using stochastic equations\cite{WN}
\begin{equation}
d\xi_t^{j,k}=-(\gamma_{j,k}+i\omega_{j,k})\xi_t^{j,k} dt +\sqrt{2\gamma_{j,k}A_{j,k}}dW_t^{j,k}\label{SPR}
\end{equation}
which are integrated from $-\infty$ to $t$\cite{WN}. Here $W_t^{j,k}$ are complex Wiener processes with properties $M[dW_t^{j,k}dW_t^{j,k}]=0$ and $M[dW_t^{j,k*}dW_s^{l,m}]=\delta_{t,s}\delta_{j,l}\delta_{k,m}$. Accurate and efficient methods for solving sets of equations like (\ref{EEs}) and (\ref{SPR}) are well established\cite{SDE,COMM}. Similar considerations apply in the case of finite temperature.

\section{Three-level model for $^{24}{\rm Mg}^+$}

The six levels which comprise the relevant electronic states of $^{24}$Mg$^+$ can be mapped to a three-level
system\cite{Jump} along the lines considered by Hulet and Wineland\cite{Master2}. To do this we defines levels
labeled 1, 2 (bright states) and 3 (dark state) from the six levels of \cite{Master2} via the correspondences
\begin{eqnarray}
\rho_{11} & \leftarrow &  \rho_{11}+\rho_{55}\nonumber \\
\rho_{22} &\leftarrow &   \rho_{33}\nonumber \\
\rho_{33} &\leftarrow &   \rho_{22}+\rho_{44}+\rho_{66}.\nonumber 
\end{eqnarray}
State 1 is now the ground state and 2 is an excited state lying about 4.4 eV above 1. The 1-2 transition is resonantly driven by a laser
and fluorescence from the 2 state is monitored by photodetectors.
Occasionally interactions with the ambient radiation field cause transitions to state 3 which is near resonant with 1. During these periods
fluorescence stops.

To model this dynamics with NMQSD we must obtain the Hulet-Wineland\cite{Master2} equations in the Markovian limit. Near steady-state the mapping
\begin{eqnarray}
R_-\rho_{33} & \leftarrow &  (2/3)\gamma \rho_{44}\nonumber \\
R_+(\rho_{11}+\rho_{22}) & \leftarrow &  (2/3)\gamma \rho_{55}\nonumber 
\end{eqnarray}
is valid where $R_-$ and $R_+$ are rates into and out of the bright manifold (1+2). These quantities 
are defined in \cite{Master2} as $R_-=8\Omega^2\gamma/9\alpha^2$ and $R_+=\Omega^2\gamma/18\alpha^2$.
Here $\Omega$ is the Rabi frequency of the laser and $\alpha$ is a Zeeman shift, while $\gamma$ is 
the spontaneous decay rate out of level 2.
These developments together with Eqs. (1) from \cite{Master2} indicate that the equations for 
the diagonal density matrix elements of the three-level system are 
\begin{eqnarray}
\dot{\rho}_{11}&=&\Omega {\rm Im} \rho_{21}+\gamma \rho_{22}+R_-\rho_{33}-R_+(\rho_{11}+\rho_{22})\nonumber \\
\dot{\rho}_{22}&=&-\Omega {\rm Im} \rho_{21}-\gamma \rho_{22}\nonumber \\
\dot{\rho}_{33}&=& -R_-\rho_{33}+R_+(\rho_{11}+\rho_{22})\nonumber.\label{HW}
\end{eqnarray}
This is the set of equations we need to reproduce as closely as possible in the Markovian limit.

A previous application of Markovian quantum state diffusion to $^{24}$Mg$^+$ introduced a set of coupling operators\cite{Jump}
\begin{eqnarray}
\hat{L}_1&=&\lambda_{12}|1\rangle\langle 2|\\
\hat{L}_2&=&\lambda_{13}|1\rangle\langle 3|\\
\hat{L}_3&=&\lambda_{31}|3\rangle\langle 1|\\
\hat{L}_4&=&\lambda(|1\rangle\langle 1|-|3\rangle\langle 3|)
\end{eqnarray}
with system Hamiltonian
\begin{equation}
\hat{H}=\frac{\hbar\Omega}{2}(|1\rangle \langle 2|+|2\rangle \langle 1|).
\end{equation}
The parameters $\lambda_{12}$, $\lambda_{13}$, $\lambda_{31}$ and $\lambda$ were treated as free variables and they were chosen
to fit experimental data. This previous study cannot therefore be considered predictive. Here we will try to choose the parameters to reproduce
the results of Hulet and Wineland\cite{Master2}.
The coupling operator $\hat{L}_1$ governs spontaneous emission from 2, $\hat{L}_2$ and $\hat{L}_3$ mediate transfers into and out of the dark state 3, and $\hat{L}_4$ models the photo-detectors\cite{Master2}.
In the Markovian limit we then obtain the Lindblad-Kossakowski\cite{CPDS} type equation
\begin{eqnarray}
\frac{d\hat{\rho}}{dt}=-(i/\hbar)[\hat{H},\hat{\rho}]+\tau\sum_{k=1}^4[\hat{L}_k\hat{\rho}\hat{L}_k^{\dag}-(1/2)\hat{L}_k^{\dag}\hat{L}_k \hat{\rho}-(1/2)\hat{\rho}\hat{L}_k^{\dag}\hat{L}_k]\label{MQSD}
\end{eqnarray}
where we have assumed that the coupling operators share a common memory time $\tau=\int_0^{\infty}dt~{\rm Re}~\alpha(t,0)$. The diagonal matrix elements then satisfy
\begin{eqnarray}
\dot{\rho}_{11}&=&\Omega {\rm Im} \rho_{21}+\tau\lambda_{12}^2\rho_{22}+\tau\lambda_{13}^2\rho_{33}-\tau\lambda_{31}^2\rho_{11}\nonumber \\
\dot{\rho}_{22}&=&-\Omega {\rm Im} \rho_{21}-\tau \lambda_{12}^2 \rho_{22}\nonumber \\
\dot{\rho}_{33}&=& -\tau\lambda_{13}^2\rho_{33}+\tau\lambda_{31}^2\rho_{11}
\end{eqnarray}
which are quite similar to Eqs. (\ref{HW}). The correspondence is not exact because Eq. (\ref{MQSD}) is positivity preserving while (\ref{HW}) is not. In any case we can now compare the two sets of equations and deduce that
\begin{eqnarray}
\lambda_{12}&=&\sqrt{\gamma/\tau}\nonumber \\
\lambda_{13}&=&\sqrt{R_-/\tau}\nonumber \\
\lambda_{31}&=&\sqrt{R_+/\tau}\nonumber \\
\end{eqnarray}
which reduces the set of unknowns in the model. Some of the remaining constants are known. For example, $\gamma= (2\pi) ~43$ MHz, $\alpha=(4\pi)~26.1 $ GHz for a magnetic field of 1.4 T. In time units of $10^{-3}\gamma^{-1}=3.7$ psec we find $\alpha =12.1$, and we set $\Omega=2$. 

The remaining unknowns are the parameter $\lambda$ associated with the photodetector, and the temperature and distribution of frequencies for the memory
function. None of these quantities can even be estimated. Accordingly we arbitrarily set $\lambda=.22/\sqrt{\tau}$ and choose a memory function common to all coupling operators at 0 K which is of the form (\ref{MEMORY}) with parameters as given in the table. This memory function is strongly non-Markovian with an initial fall-off at around $50$ time units. The parameter $\tau$ can be calculated from $\alpha(t,0)$ and we found $\tau=508.6$.

\begin{center}
\begin{tabular}{|c|c|c|}
\hline
$A_j$ & $\gamma_j$ & $ \omega_j$ \\ \hline
  2.46740754  & 0.00437729384 & -0.0934233663\\
  5.52627445  & 0.010808938 & -0.0766453125\\
  10.  & 0.0271137624  & 0.00120546934 \\
  9.58905445  & 0.0205613891 & -0.0457549602\\
  8.16208273  & 0.0269287619 &  0.0599005113\\
\hline
\end{tabular}
\end{center}

\section{Results}

We used the reformulated version of NMQSD and the Markovian QSD theory to calculate individual realizations of the stochastic dynamics
of the driven ion. In each case we chose $|\psi_0\rangle=|1\rangle$. The dynamical variable we chose to monitor was the probability
of the system to be in the bright manifold 
\begin{equation}
P(t)=|\langle 1|\psi_t\rangle|^2+|\langle 2|\psi_t\rangle|^2
\end{equation}
which is analogous to the experimentally observed fluorescence intensity. A few example trajectories are shown in Fig. 1. Note that
the dark periods tend to be more frequent and shorter for the Markovian dynamics. The Markovian trajectories are also more noisy as
one would expect.
\begin{figure*}
\subfigure[]{
\includegraphics[width=3.in,height=2.in]{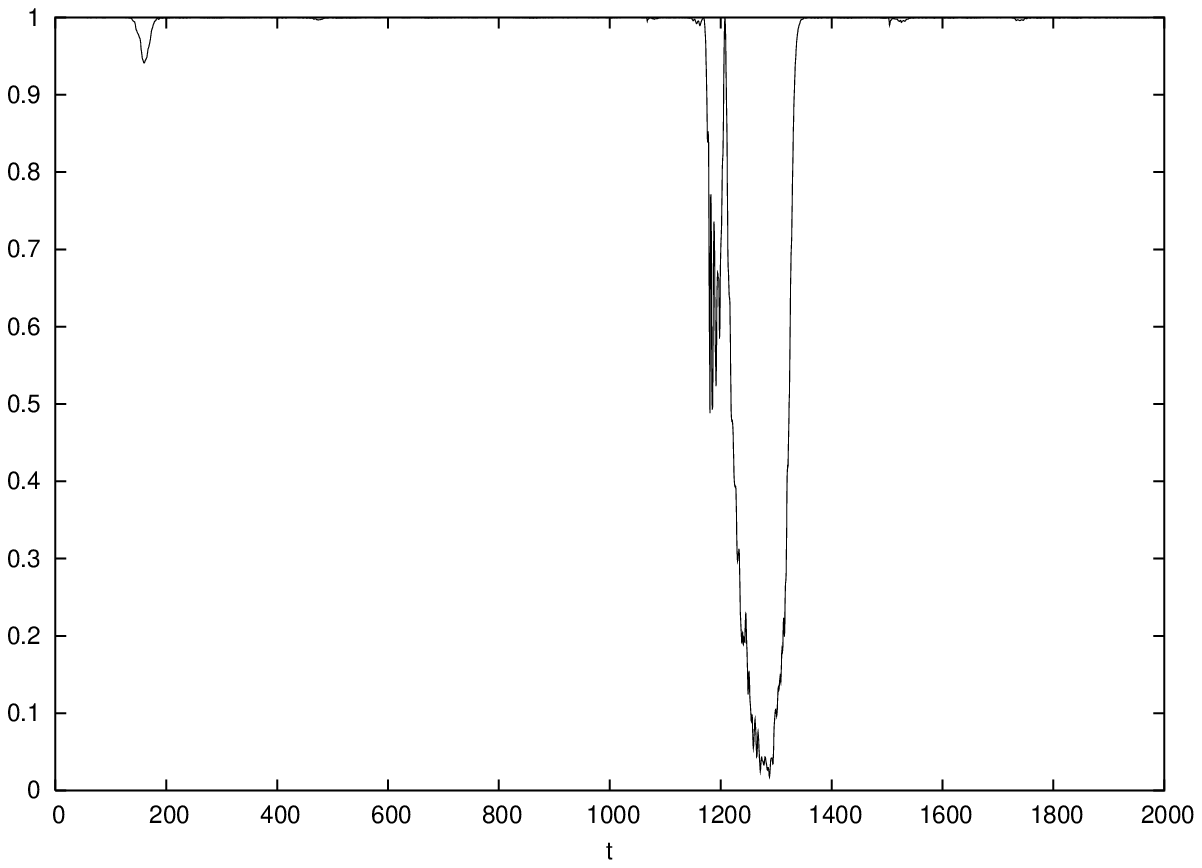}}
\subfigure[]{
\includegraphics[width=3.in,height=2.in]{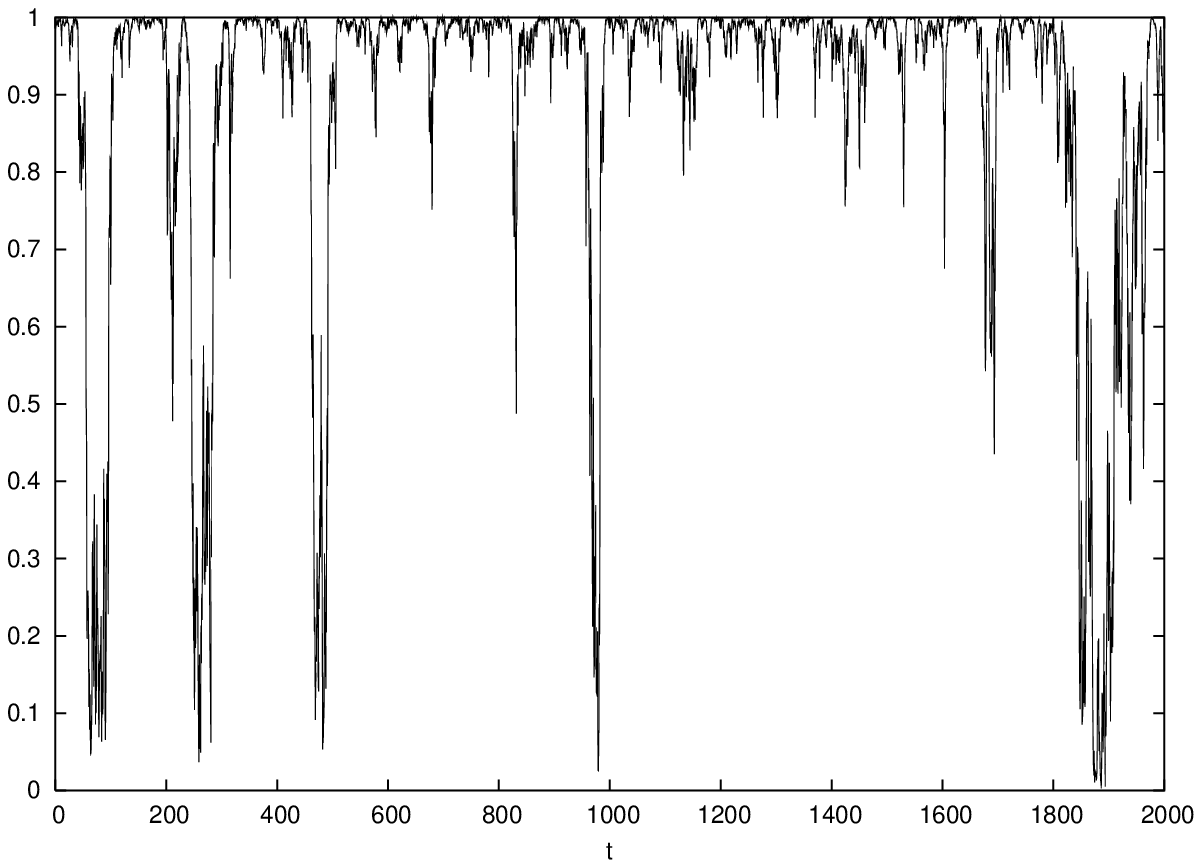}}
\subfigure[]{
\includegraphics[width=3.in,height=2.in]{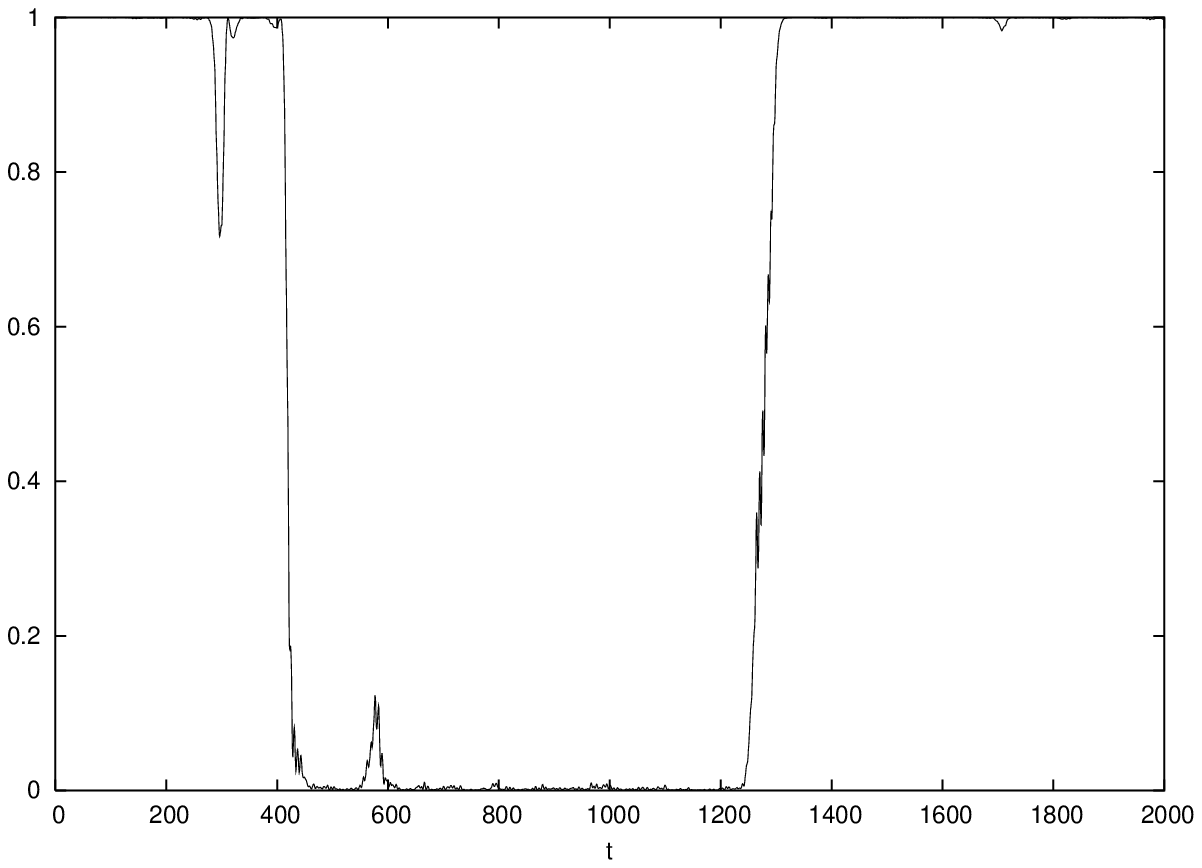}}
\subfigure[]{
\includegraphics[width=3.in,height=2.in]{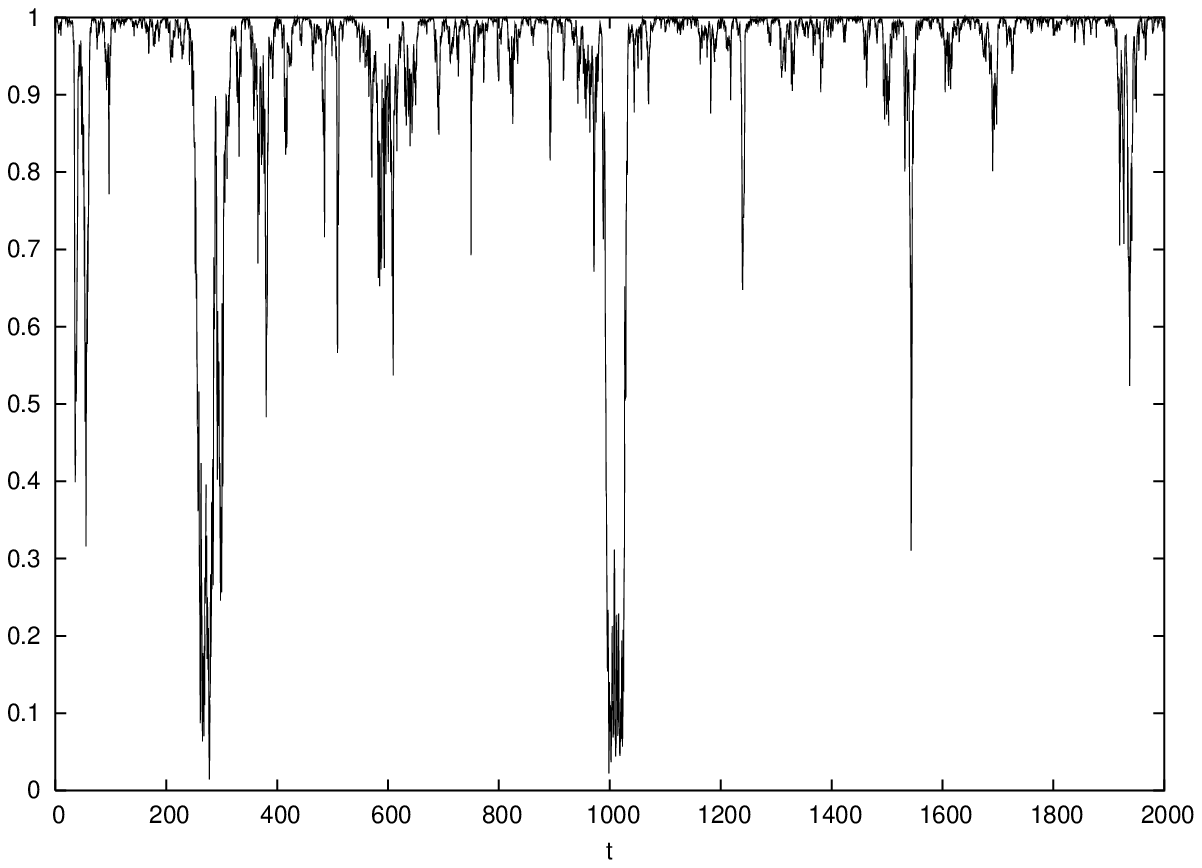}}
\subfigure[]{
\includegraphics[width=3.in,height=2.in]{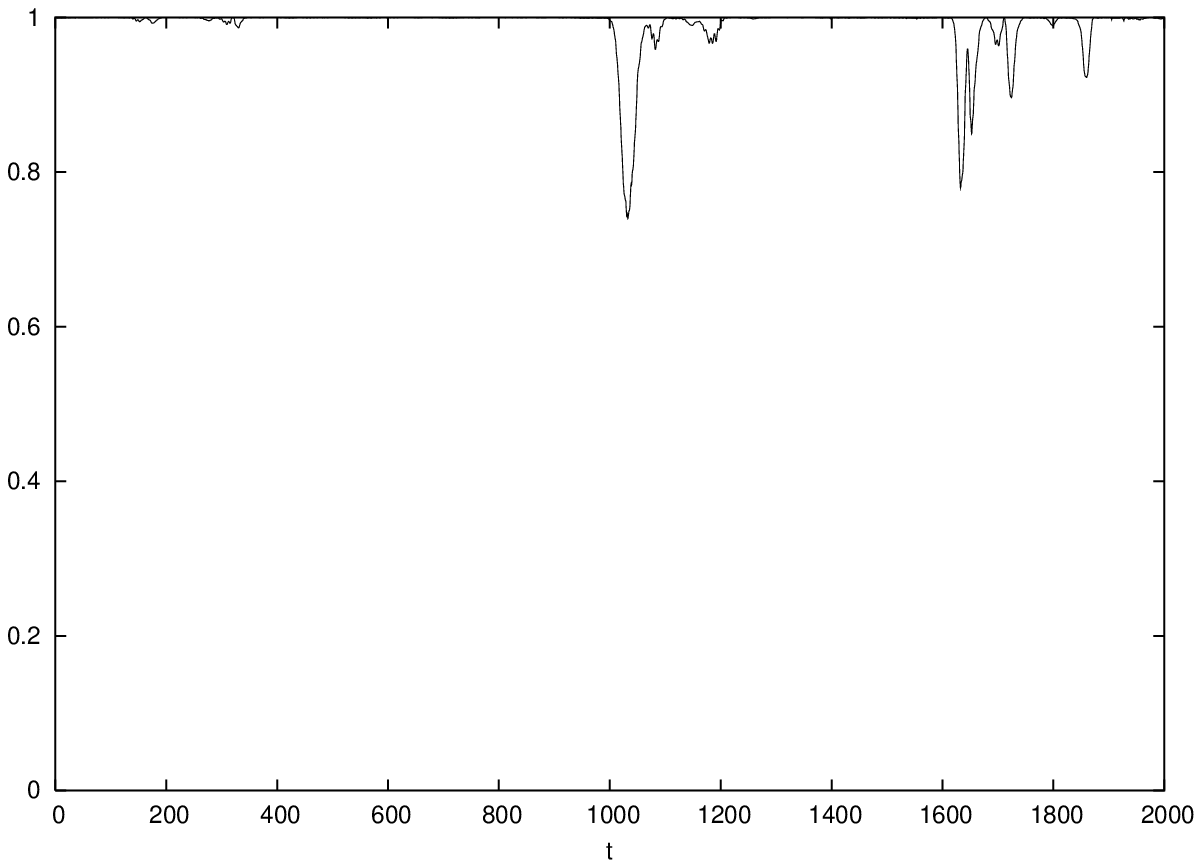}}
\subfigure[]{
\includegraphics[width=3.in,height=2.in]{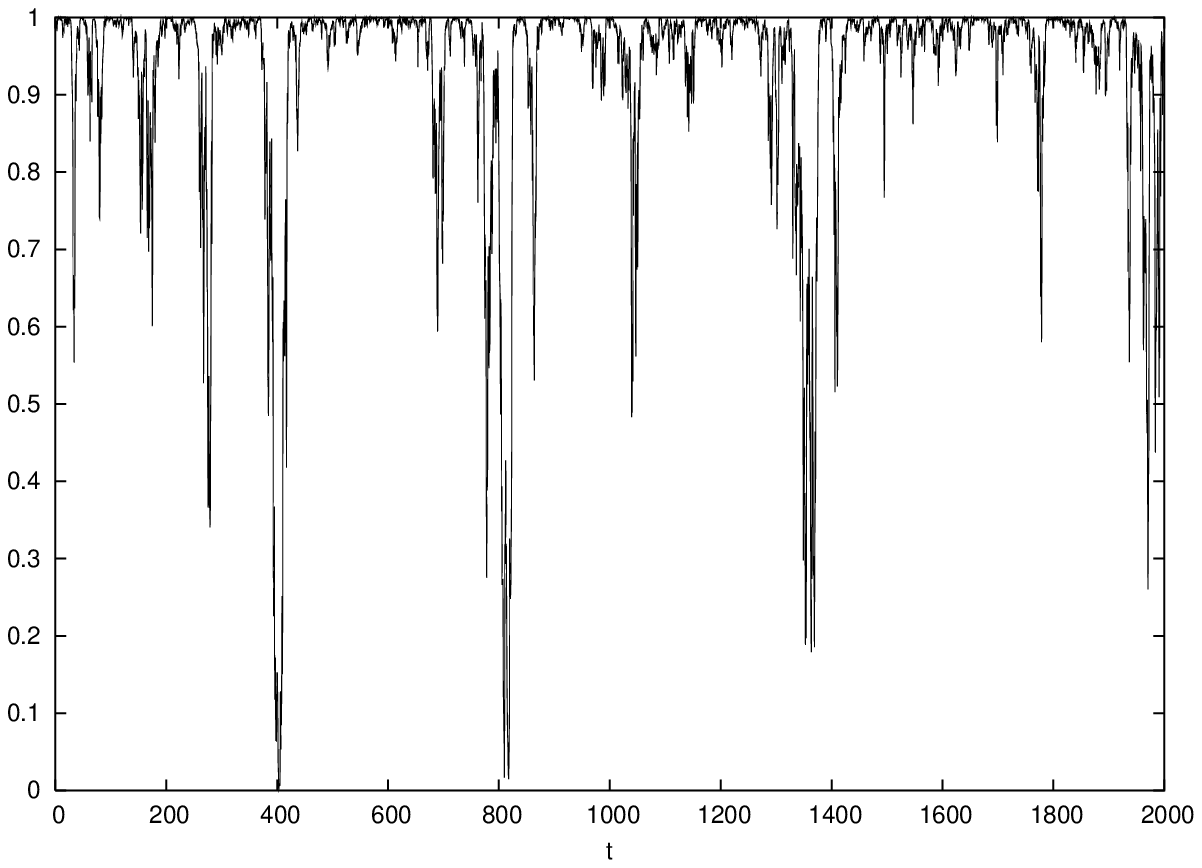}}
\caption{$P(t)$ vs $t$ for individual non-Markovian ((a), (c), (e)) and Markovian ((b), (d), (f)) trajectories}
\end{figure*}

It is customary in such jump experiments to construct a histogram of the frequency with which fluorescence intensities are
observed. Mathematically the histogram $\chi(P)$ vs $P$ is given as a limit
\begin{equation}
\chi(P)=\lim_{\Delta P\rightarrow 0}\langle (1/T)\int_0^Tdt \int_{P}^{P+\Delta P}dy ~\delta (y-P(t))\rangle
\end{equation}
which we approximate by choosing a finite but small $\Delta P$. The angle brackets denote an ensemble average over individual 
trajectories (4000 in the non-Markovian case and 10000 in the Markovian case). For $^{24}$Mg$^+$ this histogram was observed
to have two peaks, one near zero signal strength and one near maximum signal strength. The ratio of the area under the maximum 
signal peak to that under the minimum signal peak is 16 according to both theory and experiment. The numerical results we 
obtained are shown in Fig. 2.
\begin{figure*}
\subfigure[]{
\includegraphics[width=5.in,height=2.5in]{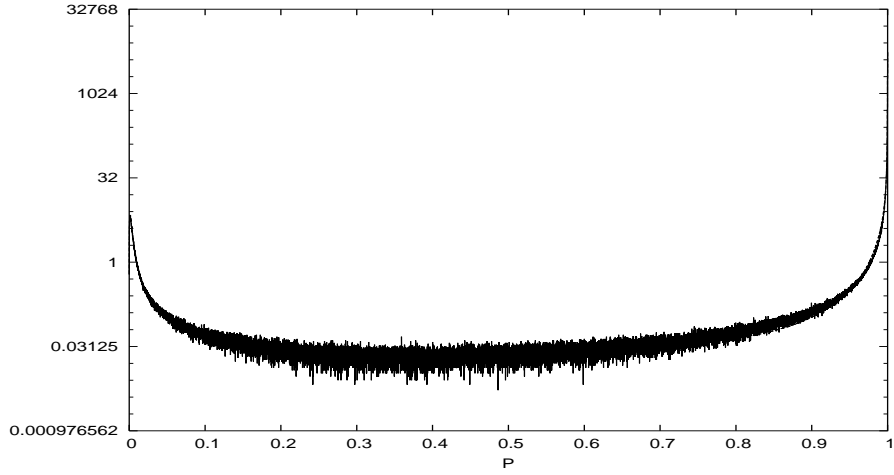}}
\subfigure[]{
\includegraphics[width=5.in,height=2.5in]{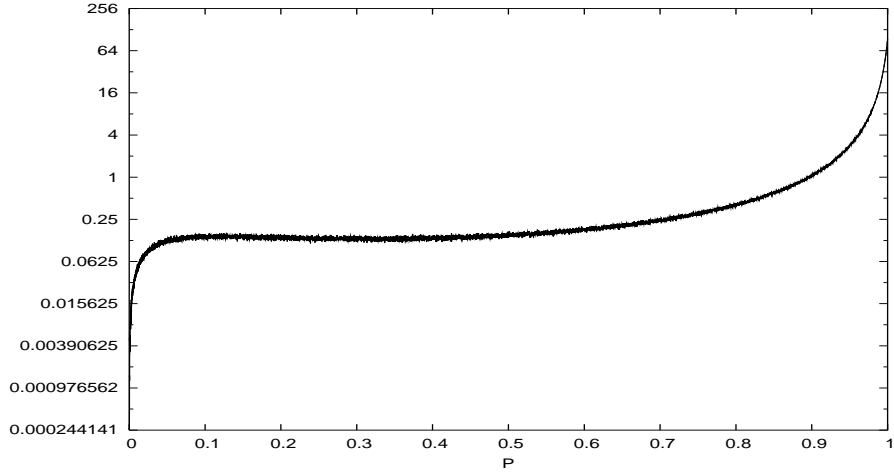}}
\caption{Non-Markovian (a) and Markovian (b) signal frequency $\chi(P)$ vs $P$ }
\end{figure*}
In the non-Markovian case two clear peaks are observed, one near $P=0$ and one near $P=1$, corresponding to occupation of the dark 
and bright manifolds respectively. To extract a ratio of areas under the peaks we fit the data to
\begin{equation}
\chi(P)=\chi_1(P)+\chi_2(P)+\chi_{background}(P)
\end{equation}
where
\begin{eqnarray}
\chi_1(P)=h_1\sqrt{P}/\{(P/w_1)^2+1\}
\end{eqnarray}
models the low signal peak and
\begin{eqnarray}
&&\chi_2(P)=h_2P/\{[(1-P)/w_2]^3+[(1-P)/w_3]^2+(1-P)/w_4+1\}\nonumber \\
&&+h_3P/\{[(1-P)/w_5]^2+(1-P)/w_6+1\}
\end{eqnarray}
models the high signal peak. The background in between was fitted to
\begin{eqnarray}
\chi_{background}(P)=h_3\{{\rm erf}[(P-P_1)/w_7]-{\rm erf}[(P-P_2)/w_8]\}.
\end{eqnarray}
Using this fitting function we were able to extract a ratio of 16.1 for the area under the strong signal peak to the area under the
weak signal peak. The best fit is shown in Fig. 3.
\begin{figure*}
\subfigure[]{
\includegraphics[width=5.in,height=2.5in]{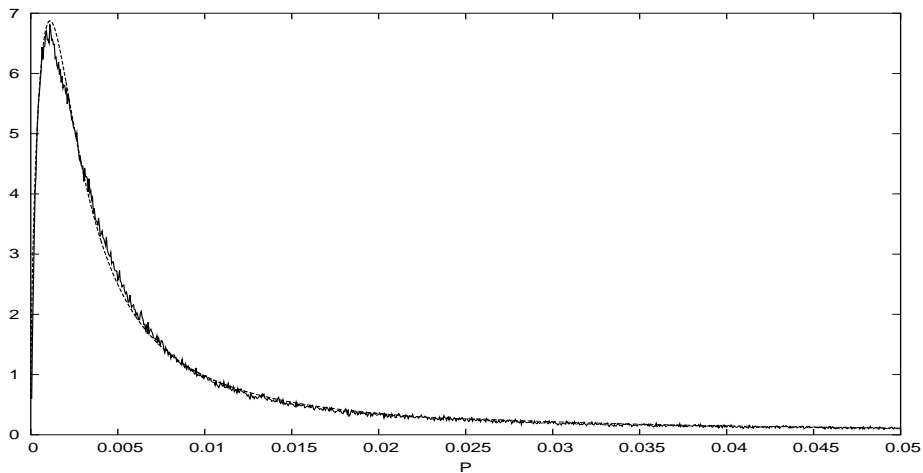}}
\subfigure[]{
\includegraphics[width=5.in,height=2.5in]{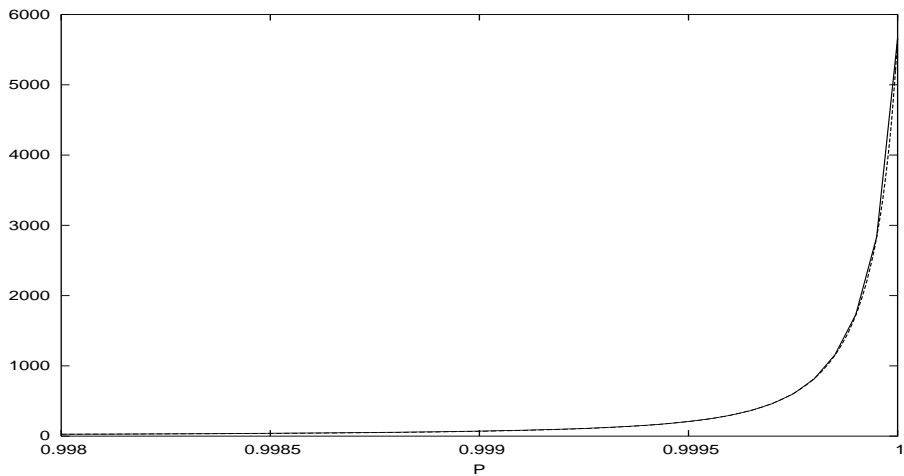}}
\caption{Non-Markovian signal frequency $\chi(P)$ vs $P$ at small and large signal values (solid curve) and best fit (dashed curve)}
\end{figure*}

In the Markovian case the existence of the low signal peak is less clear. We chose fitting functions of the form
\begin{eqnarray}
\chi_1(P)=h_1P/\{(P/w_1)^2+P/w_2+1\}
\end{eqnarray}
for the low signal peak and
\begin{eqnarray}
&&\chi_2(P)=h_2P/\{[(1-P)/w_3]^2+(1-P)/w_4+1\}\nonumber \\
&&+h_3P/\{[(1-P)/w_5]^2+(1-P)/w_6+1\}
\end{eqnarray}
for the high signal peak, which are similar to those of the non-Markovian case. The background model function was the same as in the non-Markovian case.
The ratio of areas extracted was 16.8. This minor discrepancy is likely due to lack of Monte-Carlo convergence. 

Thus, we are able to verify the ratio of areas in both the Markovian and non-Markovian cases. The mathematical form of the histogram seems
to vary little from the Markovian to non-Markovian cases even though the parameters and appearance of the two distributions are quite
different. This is encouraging since it may be possible in future to obtain time resolved experimental histograms which could be compared 
to theory to see whether there is agreement of mathematical forms. Currently, lack of time resolution and poor detection efficiency
in the experiments prevent detailed comparisons.

\section{Summary}

We have shown in considerable detail that NMQSD can be reformulated in terms of a solvable stochastic integrodifferential equation.
The reformulated theory has been tested against exact results for a number of problems, and employed to investigate a number of
problems for which exact solutions are not known\cite{WN}. We expect that the method will prove useful for many few-level-system problems
in quantum optics.

In this manuscript we applied the theory to the problem of intermittent fluorescence in $^{24}$Mg$^+$. Previous applications
 of Markovian quantum state diffusion to quantum jumps in $^{24}$Mg$^+$ treated all parameters as free variables\cite{Jump}, and hence 
cannot be considered predictive. Our more careful study shows that quantum jumps 
do indeed occur on picosecond timescales as had been speculated earlier\cite{Jump}. We also computed the probability distribution function for signal intensity and obtained the characteristic two peaks corresponding to fluorescence on and off. Our results show that both Markovian and non-Markovian versions of NMQSD reproduce the experimental and theoretical result that the ratio of the area under the bright peak is 16 times that of the area under the dark peak. We found that while the Markovian and non-Markovian distribution functions looked qualitatively different, they in fact share a very similar mathematical form. This raises a number of interesting possibilities. First, if time resolved experiments are possible it may be possible to verify the shape the histogram experimentally. This would be a much stronger result than the simple ratio of areas. Secondly, it raises the possibility of using the jump experiment to measure properties of the radiation field, since the qualitative shape of the histogram is sensitive to the memory function which contains quite a lot of information about the distribution of frequencies and temperature. 

The authors acknowledge the support of the Natural Sciences and 
Engineering Research Council of Canada.


\begin{thebibliography}{99}
\bibitem{Cnoise} We use the simpler $z_t$ notation rather than the physically suggestive $z_t^*$\cite{Wise}.
\end{thebibliography}
\end{document}